\documentclass[aps,reprint,superscriptaddress]{revtex4-1}  
\usepackage{graphicx}  
\usepackage{dcolumn}   
\usepackage{bm}        
\usepackage{amssymb}   
\usepackage{amsmath}
\usepackage{color}
\usepackage{times}

\begin{document}

\title{Statistical Physics of Design}
\author{Andrei A.~Klishin}
\affiliation{Department of Physics, University of Michigan, Ann Arbor, Michigan 48109, USA}
\affiliation{Center for the Study of Complex Systems, University of Michigan, Ann Arbor, Michigan 48109, USA}
\author{Colin P.F.~Shields}
\affiliation{Naval Architecture and Marine Engineering, University of Michigan, Ann Arbor, Michigan 48109, USA}
\author{David J.~Singer}
\affiliation{Naval Architecture and Marine Engineering, University of Michigan, Ann Arbor, Michigan 48109, USA}
\author{Greg van Anders}
\affiliation{Department of Physics, University of Michigan, Ann Arbor, Michigan 48109, 
USA}
\affiliation{Center for the Study of Complex Systems, University of Michigan, Ann Arbor, 
Michigan 48109, USA}
\email{grva@umich.edu}

\date{\today}

\begin{abstract}
  A key challenge in complex design problems that permeate science and
  engineering is the need to balance design objectives for specific design
  elements or subsystems with global system objectives. Global objectives
  give rise to competing design pressures, whose effects can be difficult to
  trace in subsystem design. Here, using examples from arrangement problems, we
  show that the systems-level application of statistical physics principles,
  which we term ``systems physics'', provides a detailed characterization of
  subsystem design in terms of the concepts of stress and strain 
from materials physics. We analyze instances of routing problems in naval
  architectures, and show that systems physics provides a direct means of
  classifying architecture types, and quantifying trade-offs between subsystem-
  and overall performance. Our approach generalizes straightforwardly to design
  problems in a wide range of other disciplines that require concrete
  understanding of how the pressure to meet overall design
  objectives drives the outcomes for component subsystems.
\end{abstract}

\maketitle

\section{Introduction}

Designing products with an emergent, overall function that is more than the sum
of their parts is a crucial challenge in science and
engineering.\cite{blanchard1990systems} Meeting this challenge is complicated by
the fact that, for many complex
products,\cite{2004toyota,morgan2006toyota,bernstein1998,singersbd,chalfant2015}
different subsystems employ diverse technologies and are designed using a
variety of methodologies.  Moreover, meeting the overall design goal for a
specific product is seldom achieved by optimal performance for every individual
subsystem.\cite{evans1959} The need to design subsystems that achieve target
performance and contribute to overall system outcomes is becoming more 
pressing.\cite{mckinseymfg,dontmakeem}
The increased pressure arises because engineered products in a wide variety of
industries now incorporate several distinct, but interconnected types of
functionality.\cite{mckinseymfg} As a result, for many modern engineered
products more economic value is added in designing a product than in
manufacturing it.\cite{dontmakeem} Making design more effective requires the
ability to understand and quantify how the design of a subsystem is affected by
overall design objectives, and how deviations from optimal performance are
affected by interaction with other subsystems.

\begin{figure}
  \includegraphics[width=0.45\textwidth]{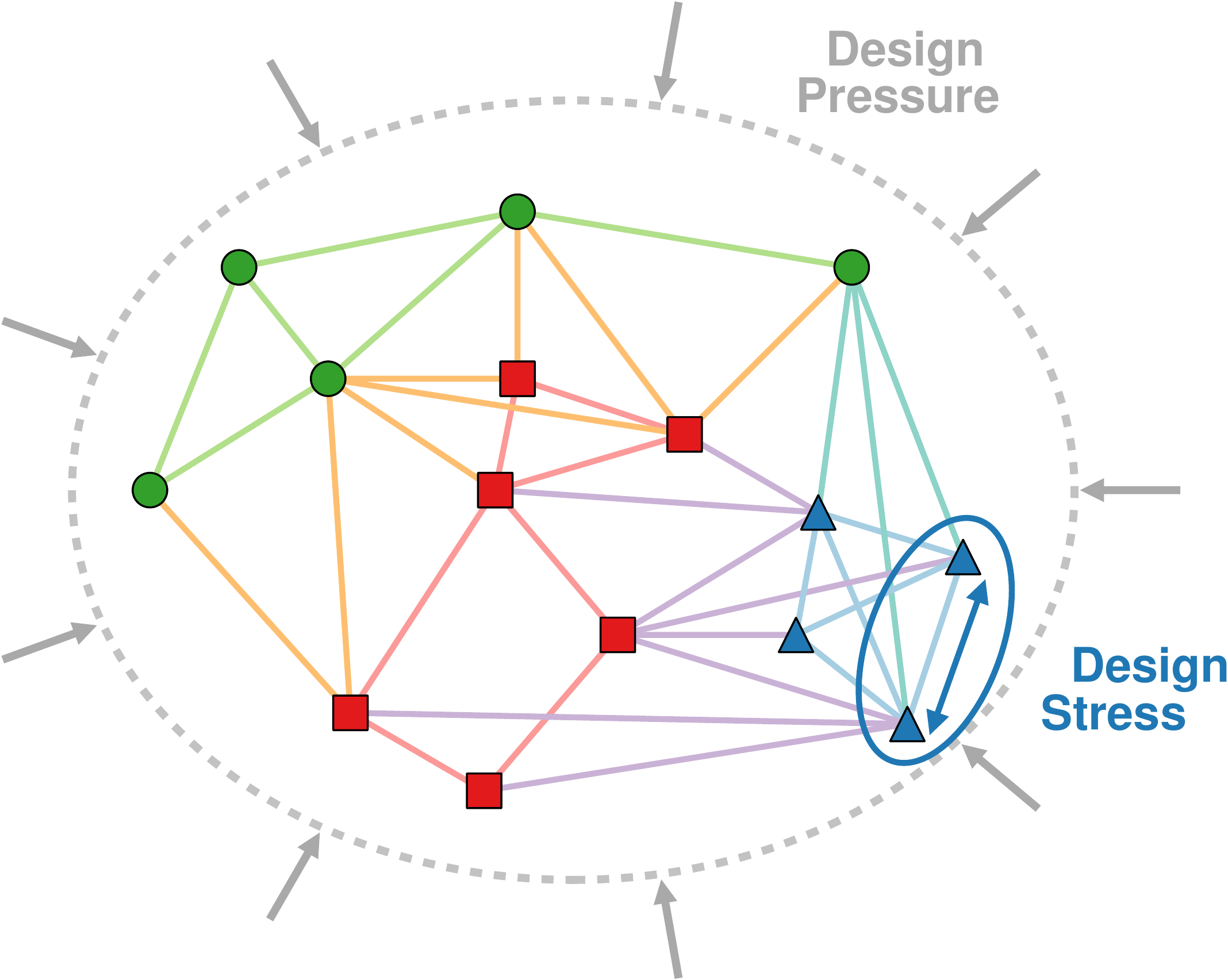}
  \caption{\textbf{Schematic of the relationship between global design pressure and 
local 
design stress in a generic design problem.} A complex system (whole network) is 
divided into 
three subsystems (represented by green, red, and blue nodes). Design pressure is represented by the 
inward pointing grey arrows and applies to all parts of the systemb. 
Locally, the global design pressure manifests itself as design stress, here 
between two blue nodes.
    \label{fig:subsystem_schematic}
  }
\end{figure}
Here, we use techniques from information theory and statistical mechanics to
show that subsystem performance and interactions can be concretely cast in terms
of ``stress'' and ``strain'' from materials physics. We illustrate this behavior
in design problems that can be cast as arrangement problems. Arrangement
problems arise in design in a wide range of disciplines, including at several
scales in electronics,\cite{dorneich1995global} as well as in distribution logistics 
\cite{aikens1985facility} and facility
layout.\cite{drira2007facility} Here, we focus on arrangement problems that arise in 
naval
architecture.\cite{Shields2017} Naval architecture, specifically that of warships or
other multi-use vessels, provides an ideal case for understanding the role of
subsystem behavior in complex engineering design. Ships incorporate several
competing design pressures,\cite{brown2003multiple,singersbd} they require
design specifications at several levels of detail,\cite{Andrews2012} and costs
frequently prevent prototype production.\cite{ross2004} Additionally, ship design has a 
need for design flexibility, i.e. it requires the consideration of nearly-redundant 
designs of comparable ``cost'' of the overall design objective. This type of design 
cannot be done via approaches that focus on finding individual designs, e.g.\ simulated 
annealing \cite{kirkpatrick1983}, that don't capture entropic drives in design.
We show that situating design problems in a 
more generic statistical physics framework facilitates the computation of \emph{local} 
``design stress'' that arises in subsystems from different
competing \emph{global} design pressures (see Fig.\ \ref{fig:subsystem+schematic} for
illustration). We demonstrate how global design
pressures from the remainder of a system induce sub-optimal subsystem
performance, which we quantify through Pareto frontiers computed using
effective, or Landau,\cite{goldenfeld} free energies.  Our approach draws on work
on effective interactions in soft matter systems without a clear separation of
scales \cite{likosrev,entint,epp} and on statistical mechanics based approaches
for materials design,\cite{digitalalchemy,miskinpnas} which we apply here at the level 
of systems. Using this ``systems physics'' approach, we compute free energies
for sample systems and show how the effects of competition between design
pressures drive subsystem designs into distinct classes.  We also use the same
method to show that it is possible to determine likely arrangements of functional
units, and routings between them, independently.  Our approach gives new
concrete, quantitative understanding of how competing design pressures affect
subsystem design in complex naval systems. Our approach can be
straightforwardly generalized to other classes of design problems involving
complex couplings between interconnected systems.

\section{Systems Physics Framework}

We seek a framework for studying tradeoffs in design problems. To do so, we begin from 
the fact that many classes of design problems can be cast in the form of a network of 
functional components.\cite{Shields2017,meadconway1980} Different candidate design 
realizations arise from different intrinsic properties of the functional units, the 
topology of the network of functional connections and, possibly, the spatial embedding 
of the functional network. For many real-world design problems this results in a 
combinatorially large space of
feasible design solutions.\cite{daskin2011network,meadconway1980,Shields2016} The 
structure of design space determines the form of tradeoffs between design considerations.

To study how the structure of design space encodes tradeoffs, we consider a 
combinatorially large set of feasible designs ($\{\sigma\}$) and a set of
design objectives ($\{\mathcal{O}_i\}$). A powerful approach to the design of complex 
engineering systems, known as Set-Based 
Design,\cite{2004toyota,morgan2006toyota,bernstein1998,singersbd} involves finding 
candidate sets of feasible designs, as opposed to focusing on a singular optimal 
design.\cite{evans1959} Different design objectives select different corners of the full 
design space into the candidate set. Given the full design space and a set of specified 
average outcomes for the design objectives ($\{\left<\mathcal{O}_i\right>\}$), an 
important task is to determine the probability ($p_\sigma$) that a given design $\sigma$ 
would be selected for inclusion in the set of candidate designs.

To construct a set of candidate designs with average outcomes 
$\{\left<\mathcal{O}_i\right>\}$ for
the design objectives, 
information theory \cite{shannon,jaynes1} indicates that the least-biased estimate of 
$p_\sigma$ is
given by maximizing the functional
\begin{equation} \label{entropyexpr}
  S = -\sum_\sigma p_\sigma \ln p_\sigma
  - \sum_i\lambda_i\left(\sum_\sigma p_\sigma \mathcal{O}_i(\sigma)
                         -\left<\mathcal{O}_i\right>\right) \; ,
\end{equation}
with respect to $p_\sigma$, where $\lambda_i$ are Lagrange multipliers enforcing
the constraint on candidate designs. Carrying out the maximization gives
\begin{equation} \label{pexpr}
  p_\sigma = \frac{1}{\mathcal{Z}}
  e^{- \sum_i\lambda_i\mathcal{O}_i(\sigma)} \; ,
\end{equation}
where $\mathcal{Z}$ is a normalization constant. In principle, further
algebraic manipulation could determine the $\lambda_i$ and yield a precise form
for $p_\sigma$. That form of $p_\sigma$ would answer the question of \emph{what}
designs are likely to be selected. \emph{Why} certain design classes are likely
to be selected, however, presents an equally important question. Answering this
question is important in untangling the dependence of specific design solutions
on overall design priorities. To answer the ``why?'' question, we note
that $p_\sigma$ has the form of Boltzmann weight in statistical physics. Using
the statistical physics approach takes us from Eq.\ \eqref{pexpr} to the
so-called partition function
\begin{equation} \label{Zexpr}
  \mathcal{Z} = e^{-\sum_i\lambda_i\mathcal{O}_i(\sigma)} \; ,
\end{equation}
in which each $\lambda_i$ quantifies the ``design pressure'' of meeting corresponding
design objective $\mathcal{O}_i$. By specifying how the variable design
pressure affects the determination of candidate designs, the partition
function provides a means to determine why candidate designs are candidates. To
concretely demonstrate the power of this approach for general design problems,
we use a specific problem from naval architecture. However, this approach
generalizes straightforwardly to other problem classes by appropriate selection
of candidate designs (${\sigma}$) and design objectives (${\mathcal{O}_i}$).

\section{Arrangement Problem Model}

\begin{figure}
  \includegraphics[width=0.45\textwidth]{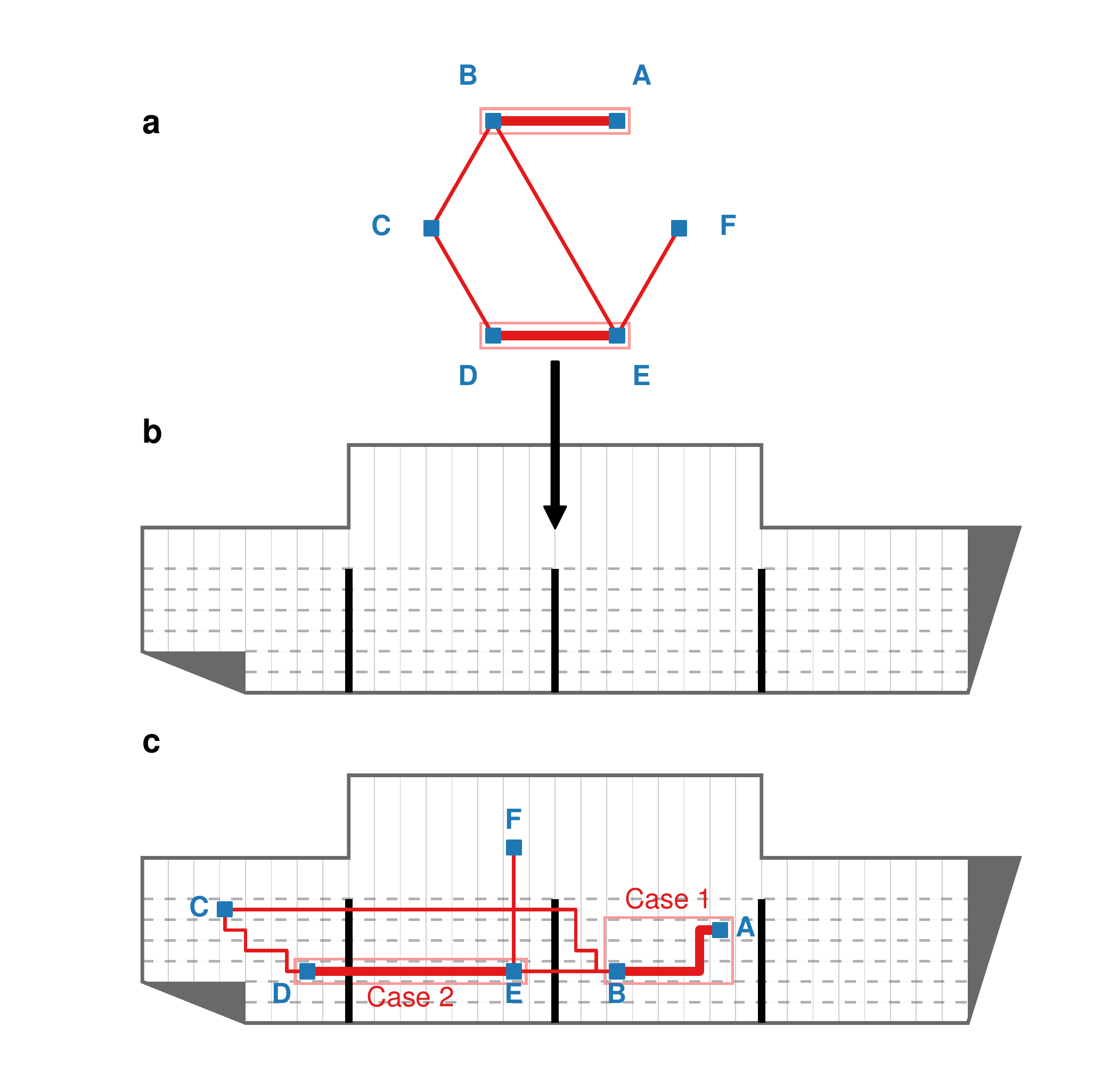}
  \caption{\textbf{Illustration of the model for arrangement problems.} The functional 
  network (\textbf{a}) is embedded into an
  inhomogeneous space (\textbf{b}), here a ship hull. Spatial embeddings
  (\textbf{c}) require
    routings between connections, with two generic cases. Case 1, routings that
    are not affected by features of the embedding space, is described in Figs.\
    2 and 3. Case 2, routings that are affected by features of the embedding
    space, here bulkheads, is described in Figs.\ 4-6.
    \label{fig:ship_cartoon}
  }
\end{figure}

We consider the spatial embedding of a subsystem of the overall functional
network that contains only two units and a single functional connection. In both
cases we choose a subsystem at random among two possible cases that differ by
whether the embedding of the remainder of the functional network localizes the
subsystem in a homogeneous space (Case 1), or a space that is structured by the
remaining ship design (Case 2). See Fig.\ \ref{fig:ship_cartoon} for an
illustration.  We show below that Case 1 exhibits behavior that results
from trade-offs between considerations of cost and flexibility, and Case 2
exhibits behavior that results from trade-offs between considerations of cost,
flexibility, and performance.

In both cases, the monetary cost expended on
routing a connection between units ($E$) is given by the ``Manhattan''
distance (the sum of horizontal and vertical steps) of a minimal path between
the units at some cost per unit length $C$. The objective for units separated by
some relative $\Delta x$ and $\Delta y$ is
\begin{equation}
  \mathcal{O}_1 \equiv E = C(\Delta x+\Delta y) \; ,
\end{equation}
and we quantify the design pressure for cost through $\lambda_1\equiv 1/T$ where $T$ is
interpreted as a ``cost tolerance''.  Low cost tolerance means that the design
pressure to minimize costs is strong, which should lead to a preference for low
cost designs.  Increasing cost tolerance weakens the design pressure to minimize
costs. Note that the limit of $T\to\infty$ represents complete indifference to
cost as a design decision factor, rather than a preference for high cost.  In
statistical physics terms, $E$ plays the role of energy, $T$ plays the role of
temperature. In addition, distinct routings and overall displacements of
the units contribute entropy, a measure of the flexibility to realize distinct
designs at fixed cost.

In addition, Case 2 models the performance penalty associated with routing
functional connections through the bulkhead. We do so with the objective
\begin{equation}
  \mathcal{O}_2 \equiv B \; ,
\end{equation}
which takes the value 1 if a routing penetrates the bulkhead and 0 if it does
not. We represent the penalty for bulkhead penetration by
$\lambda_2\equiv\gamma$.

In both cases we use statistical physics to extract design information.
Constitutive relations or ``equations of state'', evaluated via the expression
\begin{equation}\label{outcome}
  \left<O_i\right> = -\frac{\partial \ln \mathcal{Z}}{\partial \lambda_i} \; ,
\end{equation}
quantify how outcomes for design objectives are determined by design pressure.
In the specific case we consider here, fixing the design pressures through $T$ and 
$\gamma$
yields expected outcomes for $\left<E\right>$ and $\left<B\right>$, which indicate
expected costs and likelihood of bulkhead penetration, respectively.  Likewise,
the sensitivity of design outcomes to changes in design pressure is described
by ``susceptibilities'' that can be evaluated by further differentiation.
The magnitude of susceptibility is directly related to the magnitude of
fluctuations about the average design objective (see SM for more information). We
also evaluate the likely design outcomes for specific design features $S_j$
\begin{equation}
  \left<S_j\right> = \frac{1}{\mathcal{Z}}
  \sum_\sigma S_j(\sigma) e^{-\sum_i \lambda_i \mathcal{O}_i(\sigma)} \; .
\end{equation}
Finally, effective, or Landau,
free energies $F$ for different system elements (e.g.\ unit locations, routing
locations) can be computed as
\begin{equation}
  e^{-F(S_j)} \propto
  \sum_\sigma \delta(S_j(\sigma)-S_j)e^{-\sum_i \lambda_i \mathcal{O}_i(\sigma)}\;,
\end{equation}
and represent the change in the overall design objective resulting from the
competition between the design pressures. Minimal free energy corresponds to the
optimal design, whereas free energy isosurfaces represent non-optimal Pareto
frontiers. Differentiating the free energy ($-\nabla F$) yields a ``design
stress'', which quantifies how overall, global design pressure is distributed
locally among design elements in the subsystem. Similarly, ``design strain'' in
a subsystem expresses the displacement of subsystem units or routings from
optimality due to stress between subsystem and whole system design pressure.
Details of analytic and numerical computations that yield these quantities for
our model systems are described in SM.

\section{Results}
We consider the spatial embedding of a subsystem of the overall functional
network that contains only two units and a single functional connection. In both
cases we choose a subsystem at random among two possible cases that differ by
whether the embedding of the remainder of the functional network localizes the
subsystem in a homogeneous space (Case 1), or a space that is structured by the
remaining ship design (Case 2). See Fig.\ \ref{fig:ship_cartoon} for an
illustration.  We will show below that Case 1 exhibits behavior that results
from trade-offs between considerations of cost and flexibility, and Case 2
exhibits behavior that results from trade-offs between considerations of cost,
flexibility, and performance.

\subsection{Case 1, Homogeneous Embeddings: Cost/Flexibility Trade-off}
\begin{figure*}
\includegraphics[width=\textwidth]{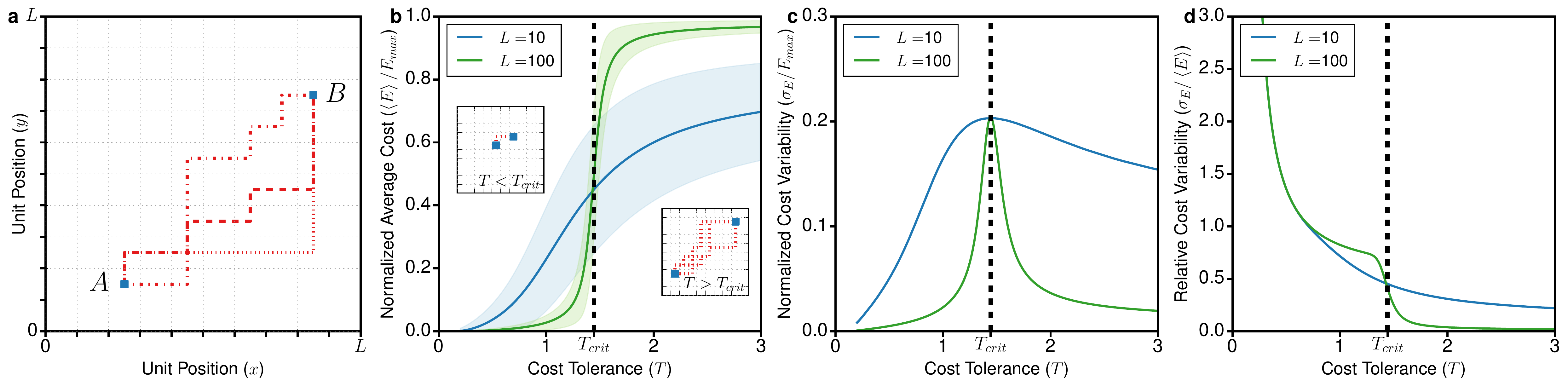}
\caption{(a) Example unit positions and routings for spatially homogeneous
  subsystem embeddings (Case 1, see Fig.\ \ref{fig:ship_cartoon}). Blue markers
  indicate unit positions, red lines indicate possible routings.
  (b) Equation of state relating cost tolerance ($T$) and average cost
  ($\left<E\right>$, in currency) normalized by maximum possible cost expended
  ($E_\textsf{max}$) for subsystems localized in an $L\times L$ region of a ship
  ($L=10$ blue curve; $L=100$ green curve). Shaded areas indicate cost
  variability. Inset images illustrate typical design realizations below
  (condensed) and above (separated) $T_\textsf{crit}=1/\ln(2)$.
  (c) Cost variability ($\sigma_E$, a susceptibility) normalized by maximum possible
  expenditure as a function of cost tolerance. The peak at $T_\textsf{crit}$ for
  a finite sized system ($L=100$) would correspond to a phase transition in the
  thermodynamic limit.
  (d) Cost variability normalized by average expenditure as a function of cost
  tolerance. Data indicate that for both large and small systems relative cost
  variability is large for low average cost designs.
  \label{fig:phase_transition_graphs}}
\end{figure*}
We consider the homogeneous embedding of a subsystem with two units, labeled $A$
and $B$, within a homogeneous region of space, here a single watertight
compartment (illustrated schematically in Fig.\ \ref{fig:ship_cartoon}c). The
location of $A$ and $B$ within the compartment, and the routing of a functional
connection between them, leads, in our model system, to a trade-off between cost
expenditure, $E$, and flexibility, measured by the routing entropy. The optimal
design of this subsystem is determined by the relative importance of cost and
flexibility, which we parametrize through the cost tolerance $T$. In Fig.\
\ref{fig:phase_transition_graphs}a we illustrate example schematic embeddings of
the subsystem of interest into a region of space of size $L\times L$. We study
examples in which the subsystem is highly localized ($L=10$) and delocalized
($L=100$) in Fig.\ \ref{fig:phase_transition_graphs}b-d. For both values of $L$
we study ensembles of design solutions at a series of values for cost
tolerance.

For $L=10$, we find that there is a slowly varying, monotonic increase in
average cost with increasing cost tolerance (Fig.\
\ref{fig:phase_transition_graphs}b, blue curve).  However, for $L=100$, where
the subsystem embedding is less constrained by the remainder of the network, we
find a sharp increase in cost around $T_\textsf{crit}=C/\ln 2$ (Fig.\
\ref{fig:phase_transition_graphs}b, green curve). This sharp increase in cost is
reminiscent of a phase transition in physical systems, and we find that the
amount of absolute cost uncertainty across feasible solutions (Fig.\
\ref{fig:phase_transition_graphs}c; akin to a susceptibility for cost) has a
peak at $T_\textsf{crit}$.  For $L=100$, when the subsystem is less constrained,
the absolute cost uncertainty is low at both low and high cost tolerance, indicating
that in those regimes routings between unit pairs are almost always cheap, or
almost always expensive relative to possible maximum cost. For $L=10$, when the
subsystem is more tightly constrained, the absolute cost uncertainty is
large over a broad range of cost tolerances.

However, when measured relative to average cost, we find that cost uncertainty is
large for both $L=10$ and $L=100$ in the limit of low cost tolerance. Fig.\
\ref{fig:phase_transition_graphs}d shows that relative cost uncertainty diverges
as cost tolerance goes to zero. This result means that even though, as expected,
low cost tolerance leads to low cost designs for the subsystem of interest,
possible design outcomes show uncertainty of 100\% or more in terms of average
cost. Though this effect might not be a large design concern if it occurred only
in the subsystem of interest, we note that our choice of subsystem was
arbitrary, so that every subsystem in the network should exhibit this effect.
A cascade of such occurrences throughout a large functional network in a complex
product, such as a ship, would lead to large macroscopic fluctuations in cost of
the overall design.

For $L=100$, Fig.\ \ref{fig:phase_transition_graphs}d indicates that as the cost
tolerance increases across the critical value, there is a sharp drop in the cost
uncertainty relative to average cost, that is driven by the sharp increase in
average cost seen in Fig.\ \ref{fig:phase_transition_graphs}b. This
indicates that above the critical cost tolerance candidate designs are high
cost, but show relatively small cost uncertainty. Taken together, the features of
the relative cost uncertainty curve indicate a fundamental trade-off: tight cost
constraints lead to wild relative cost uncertainty, whereas low relative cost
uncertainty can only be achieved at large cost.

\begin{figure*}
\includegraphics[width=\textwidth]{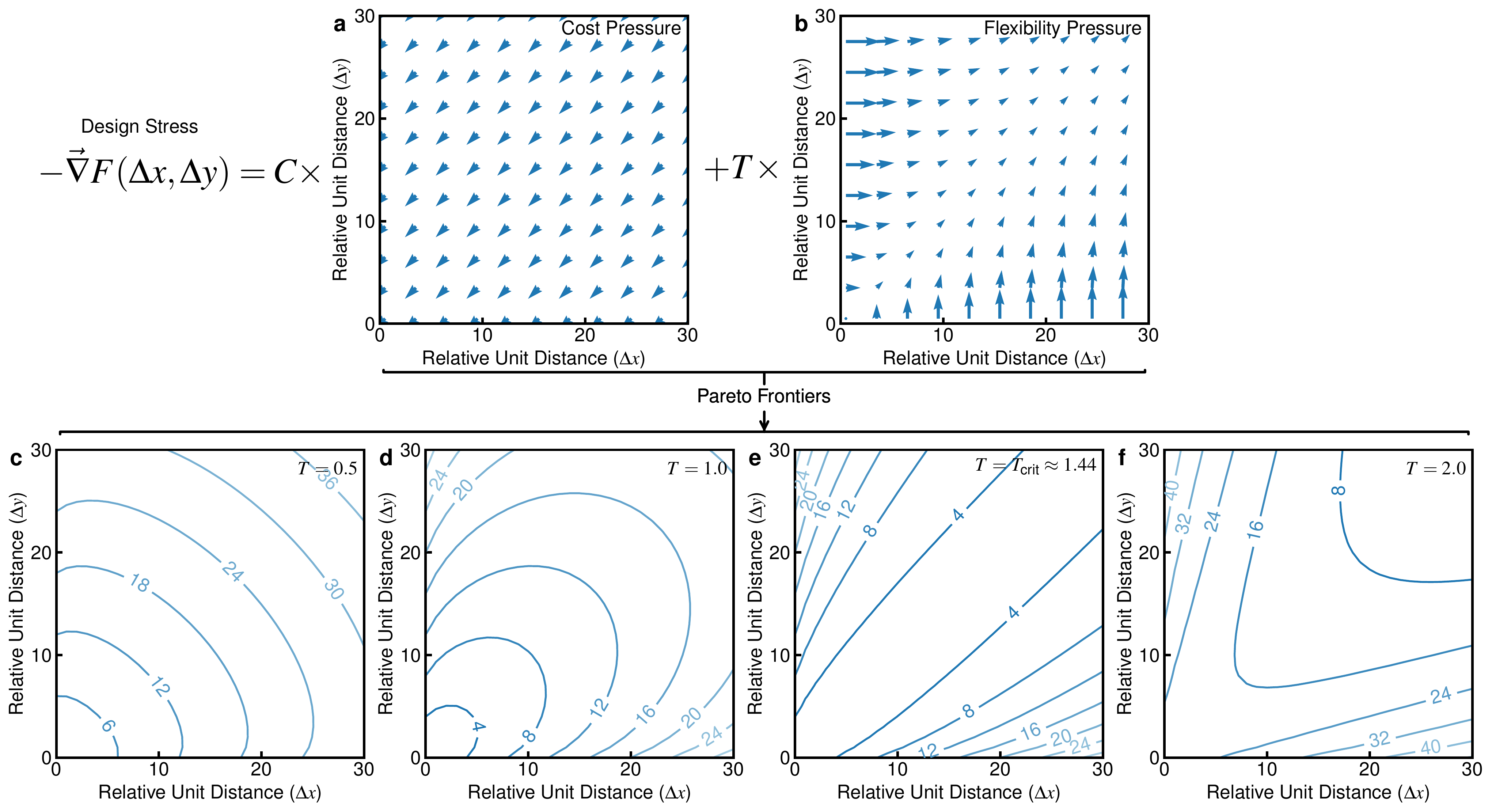}
\caption{Pressure from overall design objectives induces stress on subsystem
  design elements. For spatially homogeneous subsystem embeddings (Case 1, see
  Fig.\ \ref{fig:ship_cartoon}) design stress can be decomposed into
  contributions from cost pressure (panel a) and flexibility pressure (panel b).
  Depending on the relative strength of the design pressures, the different
  phase behaviors in Fig.\ \ref{fig:phase_transition_graphs} originate from
  underlying subsystem effects, illustrated in panels c-f. Panels c-f plot
  Pareto frontiers (Landau free energy isosurfaces) that indicate equivalent,
  suboptimal subsystem designs that could arise if the subsystem design was
  forced to sacrifice performance to the remainder of the system. At low cost
  tolerance ($T=0.5$ c; $T=1.0$ d) units are preferentially condensed. At high
  cost tolerance ($T=2.0$ f) units are preferentially separated. At the critical
  cost tolerance ($T=T_\textsf{crit}=1/\ln{2}$ there is no preferred separation
  distance.
\label{fig:contours}}
\end{figure*}
To make the origin of these behaviors more concrete, in Fig.\ \ref{fig:contours}
we fix the position of one of the units to be the origin, and examine how the
design pressures from cost (Fig.\ \ref{fig:contours}a) and flexibility (Fig.\
\ref{fig:contours}b) influence the $(x,y)$ location of the second unit. For the
case of $L=30$, we plot one quadrant, the other quadrants being related by
symmetry. Arrows indicate the relative magnitude and direction of stress that each
different form of design pressure induces on the location of the second unit.
Comparing Fig.\ \ref{fig:contours} panels a and b shows that cost and
flexibility pressures act in different directions with cost driving the units
closer together and flexibility driving them further apart. The balance between
these forces is determined by the cost tolerance, and leads to qualitatively
different outcomes depending on this value, which can be seen in the Pareto
frontiers plotted in Fig.\ \ref{fig:contours}c-f. For physics readers, we note
that Pareto frontiers correspond to isosurfaces of the Landau free energy (see,
e.g., Ref.\ \cite{goldenfeld}) for unit locations. We plot Pareto frontiers
describing the deviation from the optimal overall objective at a series of cost
tolerances. The reason for considering non-optimal solutions is that any
subsystem is only part of the overall design, and we do not expect that, in
general, overall optimal designs will correspond to optimal outcomes for all
subsystems. Non-optimal Pareto frontiers provide a means of communicating how
design pressure from the rest of the functional network could be expected to
influence the behavior of a subsystem.

When we compute the corresponding Pareto frontiers, we find that at low cost
tolerance ($T=0.1$; Fig.\ \ref{fig:contours}c), units are condensed, since the
behavior is dominated by cost minimization, which is characterized by Pareto
frontiers with constant $x+y$ in the limit of $T=0$. Increasing cost tolerance
alters the balance between cost and flexibility. Even below the critical
tolerance ($T=1.3$; Fig.\ \ref{fig:contours}d), this causes a change in shape in
the Pareto frontiers. At the critical cost tolerance ($T=T_\textsf{crit}$; Fig.\
\ref{fig:contours}e) Pareto frontiers more closely resemble surfaces
with constant $x-y$ rather than $x+y$ as we found at low temperature. Above the
critical cost tolerance ($T=2.0$; Fig.\ \ref{fig:contours}f), Pareto frontiers
reverse their order with low free energy locations for the location of the
second unit forced to the boundary.

\subsection{Case 2, Inhomogeneous Embeddings: Cost/Flexibility/Performance
Trade-offs}
We next consider the additional design pressure that arises from an
inhomogeneous embedding space. For concreteness, we represent this as a bulkhead
within the ship hull. Bulkheads are features designed to prevent water that enters
the hull through a breach from filling all parts of hull and sinking the ship.
Routings through a bulkhead are expensive and also can reduce its effectiveness, and thus 
overall ship
performance. Hence, additional performance pressure arises in the case that
elements of a subsystem are located in different bulkhead compartments
(schematic illustration in Fig.\ \ref{fig:ship_cartoon}c). We parametrize it with 
bulkhead penalty $\gamma$, acting as the second design pressure in the system. Again, 
from a large
functional network we randomly choose a subsystem comprised by a pair of units
with a single functional connection. However, we assume that the connections
between the subsystem of interest and the remainder of the functional network
drive the location of one unit to be on one side of the bulkhead and the other
unit to be on the opposite side. We allow two types of routings between the units to 
study their trade-off: one routes along the shortest path through the bulkhead and 
suffers the penalty $\gamma$; the other routes along the shortest path around the 
bulkhead, with no penalty. For concreteness we give results for systems of
fixed size ($20\times 20$ with a vertical bulkhead in the middle) which are 
representative of the general behaviors we observe.
See SM for results for other system sizes.

\begin{figure*}
\includegraphics[width=\textwidth]{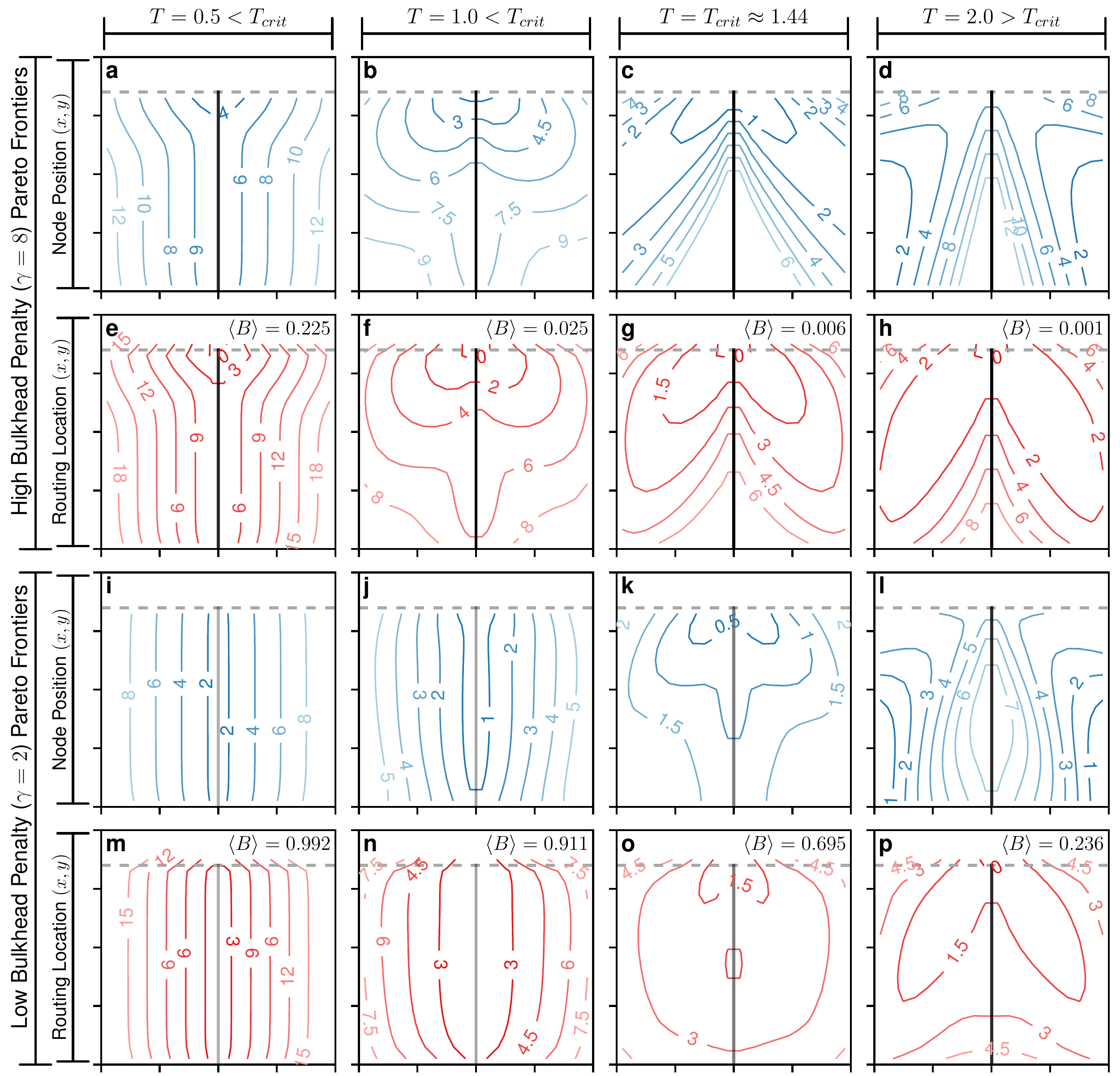}
\caption{Pareto frontiers (Landau free energy isosurfaces) for unit and routing
  locations for spatially inhomogeneous subsystem embeddings (Case 2, see Fig.\
  \ref{fig:ship_cartoon}) for different cost tolerances $T$ ($T=0.5$ first
  column, $T=1.0$ second column, $T=T_\textsf{crit}$ third column, $T=2.0$
  fourth column) and performance penalties for bulkhead penetration ($\gamma=8$
  top two rows, $\gamma=2$ bottom two rows). Blue curves indicate unit
  positions, normalized so that most favorable unit locations have value 0, with
  increasing values indicating the loss in subsystem objective in units of the
  cost tolerance. Red curves indicate routing locations, normalized so that
  locations through which connections route with absolute certainty have value
  0, and increasing values indicate the reduction in subsystem objective of
  routing through a given location in units of cost tolerance.
\label{fig:comparison_panels}}
\end{figure*}
Compared with Case 1, breaking spatial homogeneity makes the relationship
between route paths and unit locations more complicated. This complication
arises because routings now couple to both unit positions and geometric
features. Because of this, we study unit positioning and routing separately. As
in Case 1, we compute Pareto frontiers via Landau free energies, but in this
case we do so by integrating out the degrees of freedom of units and routings
separately. Fig.\ \ref{fig:comparison_panels} shows Pareto frontiers for unit
routing positions as a function of cost tolerance for bulkheads with
representative high ($\gamma=8$, panels a-h) and low ($\gamma=2$, panels i-p)
bulkhead penalty. The difference of $\Delta \gamma=6$ between the two values
implies that the relative statistical weight of routing through the bulkhead
changes roughly by a factor of $e^6\sim 400$, and the effects on node
positioning are immediately visually apparent. Also apparent is the effect of
$\Delta\gamma$ on design performance, characterized by the $\left<B\right>$,
i.e.\ the fraction of all designs that route through the bulkhead. See SM for
computation details.

At high bulkhead penalty ($\gamma=8$), and low cost tolerance ($T=0.5$) Pareto
frontiers for unit locations (Fig.\ \ref{fig:comparison_panels}a) and routing
(Fig.\ \ref{fig:comparison_panels}e) both indicate strong coupling to the top of
the bulkhead. Results for increased cost tolerance ($T=1.0$) that is still below
$T_\textsf{crit}$ indicate that unit locations are less strongly coupled to the
bulkhead (Fig.\ \ref{fig:comparison_panels}b).  Comparison with results for
routing (Fig.\ \ref{fig:comparison_panels}f) indicate that this coincides with a
drop in fraction of designs that route through the bulkhead by nearly an order
of magnitude ($\left<B\right>=0.025$ at $T=1.0$, c.f.\ $\left<B\right>=0.225$ at 
$T=0.5$), 
and
though routes remain strongly localized at the top of the barrier, Pareto
frontiers at equivalent objective cost (free energy) are further from the
bulkhead. These trends continue through $T_\textsf{crit}$ (Fig.\
\ref{fig:comparison_panels}c,g). However, above $T_\textsf{crit}$ ($T=2.0$)
Fig.\ \ref{fig:comparison_panels}d we observe that although the units delocalize
from the bulkhead (Fig.\ \ref{fig:comparison_panels}d), the routings remain strongly
coupled to the top of the bulkhead, and the probability that a design routes
through the bulkhead drops to $\left<B\right>=0.001$. Comparing unit locations (Fig.\
\ref{fig:comparison_panels}a-c) and routing locations (Fig.\
\ref{fig:comparison_panels}e-g) indicates that at or below $T_\textsf{crit}$
unit locations are correlated with routing locations. However, above
$T_\textsf{crit}$ (Fig.\ \ref{fig:comparison_panels}d,h) indicates that most
probable unit locations do not correspond to most probable routing locations.

We contrast the above results at high bulkhead penalty ($\gamma=8$, Fig.\
\ref{fig:comparison_panels}a-h) with low bulkhead penalty ($\gamma=2$, Fig.\
\ref{fig:comparison_panels}i-p). At low cost tolerance ($T=0.5$)
we see that relaxing the bulkhead penalty still causes the unit positions
to localize near the bulkhead
(Fig.\ \ref{fig:comparison_panels}e) 
but the units no longer localize near the top of the bulkhead as they did at
high bulkhead penalty (Fig.\ \ref{fig:comparison_panels}a). Likewise, routings
no longer localize near the top of the bulkhead (Fig.\
\ref{fig:comparison_panels}m), but follow the unit locations and pierce the
bulkhead with high probability ($\left<B\right>=0.992$). At increased cost tolerance 
($T=1.0, T_\textsf{crit}$) the localization at the top of the bulkhead appears again 
(Fig.\ \ref{fig:comparison_panels}j-k,n-o). At high cost tolerance $T=2.0$ the units 
again delocalize from the bulkhead (Fig.\ \ref{fig:comparison_panels}l,p) and the cases 
$\gamma=2$ and $\gamma=8$ start looking very similar.

\begin{figure*}
\includegraphics[width=\textwidth]{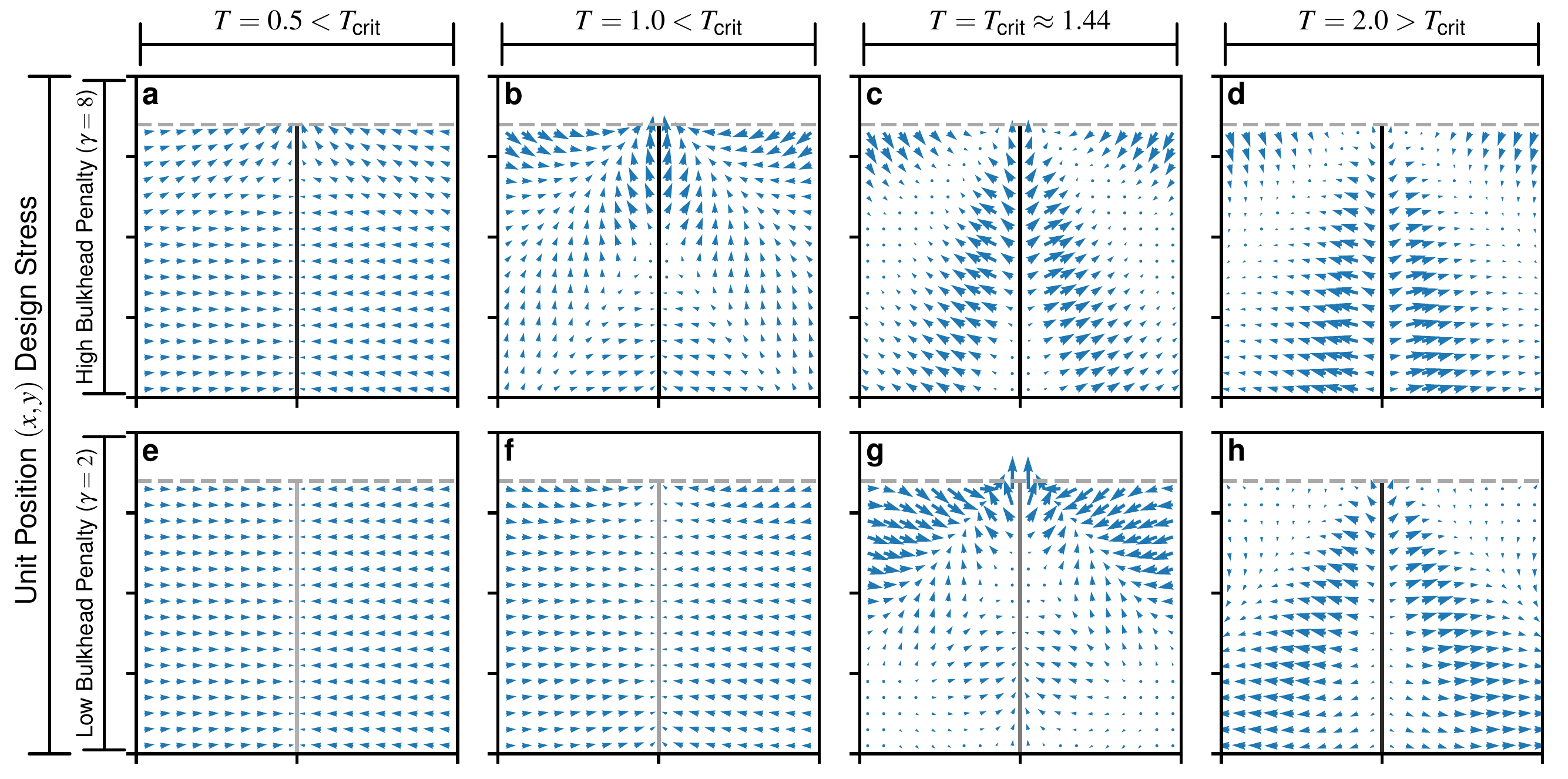}
\caption{Design stress for unit locations in spatially inhomogeneous
  subsystem embeddings (Case 2, see Fig.\ \ref{fig:ship_cartoon}) for different
  cost tolerances $T$ ($T=0.5$ first column, $T=1.0$ second column,
  $T=T_\textsf{crit}$ third column, $T=2.0$ fourth column) and performance
  penalties for bulkhead penetration ($\gamma=8$ top row, $\gamma=2$ bottom
  row). Plots indicate that if a unit was sited at the origin of an arrow in
  response to whole system design pressure, design pressure acting on the
  subsystem alone would drive the unit in the direction of the arrow, with a
  strength proportional to the length of the arrow.
\label{fig:quiver}}
\end{figure*}
To further understand the competing design pressures of cost, flexibility, and
performance, we compute design stress in unit positioning (see Fig.\
\ref{fig:quiver}). At a given unit position (corresponding to ``strain'' in the
language of materials science) design stress indicates the magnitude and
direction in which changing the placement of the unit would lead to the greatest
decrease in the overall objective cost for the subsystem. We find that at low cost
tolerance ($T=0.5$, Fig.\ \ref{fig:quiver}a,e), design stress is directed primarily
toward the bulkhead, with discernible stress toward the top of the compartment
for high cost penalty. An increase in cost tolerance ($T=1.0$, Fig.\
\ref{fig:quiver}b,f) leads to similar design stress at low bulkhead penalty
(Fig.\ \ref{fig:quiver}f) but a more intricate pattern of stress at high
bulkhead penalty (Fig.\ \ref{fig:quiver}b) that includes regions with stress
toward and away from the both the bulkhead and the top of the compartment.
Similarly, complex patterns of stress occur at both low and high bulkhead
penalty at $T_\textsf{crit}$ (Fig.\ \ref{fig:quiver}c,g). At high cost tolerance
($T=2.0$, Fig.\ \ref{fig:quiver}d,h), the pattern of design stress is
predominantly away from the bulkhead.

\begin{figure}
  \includegraphics[width=0.5\textwidth]{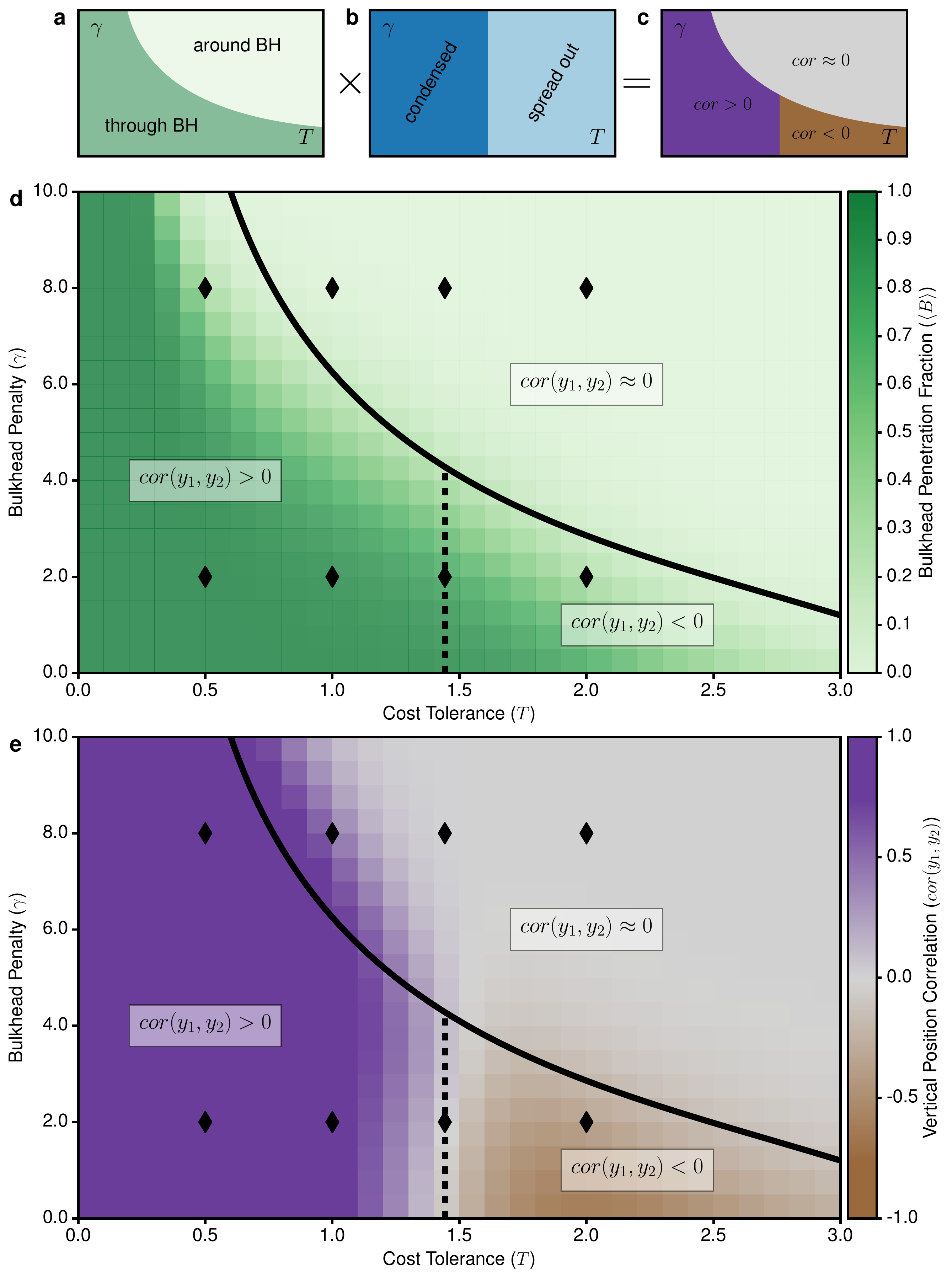}
  \caption{Phase diagram for spatially inhomogeneous subsystem embeddings (Case 2,
    see Fig.\ \ref{fig:ship_cartoon}), summarizing effects of performance
    penalties associated with bulkhead penetration ($\gamma$) and cost
    tolerance
    ($T$).
    (a) Schematic illustration of effects of $T$ and $\gamma$ on bulkhead
    penetration (performance proxy) for arbitrary subsystem localization.
    (b) Schematic illustration of effects of $T$ and $\gamma$ on unit
    separation (cost proxy) for arbitrary subsystem localization.
    (c) Schematic illustration of combination of bulkhead penetration
    (performance
    proxy) and unit separation (cost proxy) on vertical correlation in unit
    layout
    (architecture-class proxy) for arbitrary subsystem localization.
    (d) Quantitative phase plot for subsystem localization of fixed size
    $L=20$. Green shade indicates average system performance
    (bulkhead penetration probability).
    (e) Quantitative phase plot for vertical correlation in unit layout 
    (architecture-class proxy). Markers in (d-e) indicate $T,\gamma$ values
    corresponding to plots in Figs.\ \ref{fig:comparison_panels} and
    \ref{fig:quiver}.
    \label{fig:phase_diagram}}
\end{figure}
The behaviors we find that arise from the competition between cost, flexibility,
and performance design pressures can be classified qualitatively according to
the phase diagram in Fig.\ \ref{fig:phase_diagram}. In Fig.\
\ref{fig:phase_diagram} we show, schematically, the effects of bulkhead penalty $\gamma$
and cost tolerance $T$ on bulkhead penetration (a) and relative unit distance
(b). The combination of these effects also results in a complicated emergent
relationship between the vertical positions of the units (c). To provide a more concrete 
and quantitative example, panels (d) and (e) show respectively the bulkhead penetration 
fraction and the correlation in vertical node positions for the same system of size 
$L=20$.

\section{Conclusion}
We developed a general, statistical physics framework for analyzing complex
design problems. We demonstrated the application of this framework to
characterizing tradeoffs between competing design presures. For concreteness, we
studied trade-offs between competing design pressures of cost, flexibility, and
performance in arrangement problems from naval architecture design. We analyzed
ship models by applying physics principles at the systems-level and found a rich
pattern of behavior. We gave an explicit formulation of Pareto frontiers in
terms of isosurfaces of Landau free energy, and computed ``design stress'' induced
by sub-optimal subsystem embedding. Our framework recasts common design
challenges in terms of the well-understood concepts of pressure, stress and
strain. We find that these concepts, which are typically used to characterize
the behavior of materials, also provide a means of characterizing system-level
behavior.

Our approach opens new avenues for addressing design challenges that arise in
complex systems. Our framing of system design in terms of statistical mechanics
has some technical overlap with optimization approaches based on simulated
annealing.\cite{kirkpatrick1983}
Simulated annealing invokes thermodynamics by using a fictitious Hamiltonian
cooled \emph{in silico} to zero temperature to find the global minimum of an
objective function. Our approach with minimally biased probability
distributions, though derived from information theory, is mathematically
equivalent to a fictitious Hamiltonian held at a constant \emph{finite} temperature.
Maintaining finite temperature highlights the role of design pressures that arise from 
flexibility and become relevant in combinatorially large optimization
spaces, and in early stage design.\cite{chalfant2015}  We found this approach gives 
important information about the systems of
interest: the separation of subsystem designs into different architecture
classes; knowledge about where paths between units are likely to route even if
the unit locations are not specified, and vice versa; knowledge about cost
variability for low and high subsystem cost tolerance; understanding of how
different design objectives create design stress on subsystems. All of these
forms of knowledge are crucial in the early design stages of a broad class of
complex design problems.

Finally, physics concepts and principles are typically used to understand the
behavior of a part of a larger system. E.g.\ for a ship it is common to: use the
physics of electromagnetism to understand the function of a radar; use materials
physics to understand the properties of a hull; use solid state physics to
understand the properties of electronics; use hydrodynamics to
understand the interaction of a hull with water; use thermodynamics to
understand the function of an engine. Here, without explicit reference to the
underlying physical nature of the component systems, we showed that familiar
physics concepts, such as pressure, stress, and strain, via the principles of
statistical mechanics, give new insight into the architecture of a ship as a
whole. Our focus on an established,\cite{Shields2017}
minimal model of ship design was motivated both by pressing challenges in
naval architecture, and by the goal of providing a concrete, self-contained example of 
our approach. However, our ``systems
physics'' approach generalizes straightforwardly in several respects: to more
detailed models of naval architecture,
to subsystems with more units, and more complex functional connections,
and, most importantly, and to other classes of systems-level design problems.
Systems-level applications of physics have led to constructive engagements
between physics and economics,\cite{mantegna1995scaling,econophysics} network
science,\cite{AlbertBarabasi,strdynnet} and
epidemiology.\cite{pastor2001epidemic,goh2007human} We believe the present
systems-level application of physics will lead to a similar constructive
engagement with design problems in a wide variety of domains.

We thank B.\ Ames, L.\ Conway, and M.\ Newman for several helpful discussions.
This work was supported by the U.S. Office of Naval Research Grant Nos.\
N00014-17-1-2491 and N00014-15-1-2752 as well as Government support under
contract FA9550-11-C0028 and awarded by the Department of Defense, Air Force
Office of Scientific Research, National Defense Science and Engineering Graduate
(NDSEG) Fellowship, 32 CFR 168a.

\bibliography{gmaster}
\end{document}